\documentclass[
 reprint,
 superscriptaddress,
 amsmath,amssymb,
 aps,
]{revtex4-2}

\usepackage{graphicx}
\usepackage{dcolumn}
\usepackage{bm}
\usepackage{gensymb}
\usepackage{siunitx}

\begin{document}

\title{Stark Effect of Blue Quantum Emitters in Hexagonal Boron Nitride}

\author{Ivan Zhigulin}
\thanks{These authors contributed equally}
\affiliation{School of Mathematical and Physical Sciences, University of Technology Sydney, Ultimo, New South Wales 2007, Australia}
 
\author{Jake Horder}
\thanks{These authors contributed equally}
 \affiliation{School of Mathematical and Physical Sciences, University of Technology Sydney, Ultimo, New South Wales 2007, Australia}
 
\author{Victor Ivády}
 \affiliation{ Max-Planck-Institut für Physik komplexer Systeme, Nöthnitzer, D-01187 Dresden, Germany}
 \affiliation{Department of Physics, Chemistry and Biology, Linköping University, Linköping, Sweden}

\author{Simon J. U. White}
 \affiliation{School of Mathematical and Physical Sciences, University of Technology Sydney, Ultimo, New South Wales 2007, Australia}

\author{Angus Gale}
 \affiliation{School of Mathematical and Physical Sciences, University of Technology Sydney, Ultimo, New South Wales 2007, Australia}
 
\author{Chi Li}
 \affiliation{School of Mathematical and Physical Sciences, University of Technology Sydney, Ultimo, New South Wales 2007, Australia}

\author{Charlene J. Lobo}
 \affiliation{School of Mathematical and Physical Sciences, University of Technology Sydney, Ultimo, New South Wales 2007, Australia}

\author{Milos Toth}
 \affiliation{School of Mathematical and Physical Sciences, University of Technology Sydney, Ultimo, New South Wales 2007, Australia}
  \affiliation{ARC Centre of Excellence for Transformative Meta-Optical Systems, University of Technology Sydney, Ultimo, New South Wales 2007, Australia}

\author{Igor Aharonovich}
 \email{igor.aharonovich@uts.edu.au}
 \affiliation{School of Mathematical and Physical Sciences, University of Technology Sydney, Ultimo, New South Wales 2007, Australia}
 \affiliation{ARC Centre of Excellence for Transformative Meta-Optical Systems, University of Technology Sydney, Ultimo, New South Wales 2007, Australia}

\author{Mehran Kianinia}
\email{mehran.kianinia@uts.edu.au}
 \affiliation{School of Mathematical and Physical Sciences, University of Technology Sydney, Ultimo, New South Wales 2007, Australia}


\begin{abstract}
Inhomogeneous broadening is a major limitation for the application of quantum emitters in hBN to integrated quantum photonics. Here we demonstrate that so-called `blue emitters' with an emission wavelength of \SI{436}{nm} are less sensitive to electric fields than other quantum emitters in hBN. Our measurements reveal a weak, predominantly quadratic Stark effect that indicates a negligible transition dipole moment of the defect. Using these results, we discuss implications for the symmetry of the defect and use density functional theory simulations to identify a likely atomic structure of blue emitters in hBN.
\end{abstract}

\maketitle

Quantum emitters in hexagonal boron nitride (hBN) are recognized for their brightness and robustness against temperature and harsh environments \cite{tran_quantum_2016, li_integration_2021, Vogl_space_2019, Kianinia_review_2022, Luxmoore_STED_phonon_2021, Boltasseva_hBN-OnChip_2021, Hofmann_EmitterLocalization_2021}. Most recently, optical access to spin properties for some of these defects has been demonstrated at room temperature \cite{atature_material_2018, gottscholl_initialization_2020, TongcangLi_femtosecond_2021, wrachtrup_spin-vdWaals_2021, stern_room-temperature_2022}. Moreover, the van der Waals structure of the host crystals provides compelling opportunities for the development of quantum technologies based on quantum emitters in hBN \cite{Proscia_icrocavity_2020, froch_coupling_2020}. However, broad adoption of hBN quantum emitters in integrated quantum photonics has been limited by inhomogeneous broadening of the emission lines. The dominant cause of inhomogeneous broadening are fluctuating electric fields resulting from charge transitions in the vicinity of quantum emitters. The fields interact with electric dipoles of the defects and give rise to spectral diffusion. The effect has been studied extensively in various solid state systems such as NV centers in diamond \cite{ruf_optically_2019}, point defects in silicon carbide, and rare earth doped crystals \cite{zhong_nanophotonic_2015}. In the case of hBN, this interaction is very strong for some emitters, and has been utilized to realize massive Stark shifts using applied electric fields \cite{xia_room-temperature_2019, noh_stark_2018, nikolay_very_2019}. Whilst this is useful for spectral tuning of quantum emitters, the strong interaction gives rise to linewidth broadening and spectral diffusion that limit the usefulness of these emitters.

To this end, there has been work focused on eliminating spectral diffusion by surface passivation, and optical re-pumping \cite{white_optical_2020, li_nonmagnetic_2017}. Fourier-limited linewidth has been achieved for a defect in hBN by charge manipulation/depletion using heterostructure devices \cite{akbari_lifetime_2022}. An alternative approach to overcome spectral diffusion is to identify point defects with small or ideally zero permanent electric dipole moments, such as  the SiV center in silicon carbide or centrosymmetric defects such as group IV defects in diamond \cite{de_santis_investigation_2021, ruhl_stark_2020}. The absence of a permanent transition dipole in these defects results in reduced sensitivity to electric fields and small inhomogeneous linewidth broadening.
 
Recently, a group of emitters in hBN known as `blue emitters' has been identified as being spectrally stable with a narrow emission at \SI{436}{nm} \cite{shevitski_blue-light-emitting_2019, fournier_position-controlled_2021}. These defects can be created deterministically using electron beam irradiation and have been associated with the presence of a UV emission at \SI{4.1}{eV} \cite{gale_site-specific_2022}. Interestingly, the blue emitters exhibit stable single photon emission at cryogenic temperatures with spectral diffusion that is negligible compared to other quantum emitters studied to date in hBN \cite{horder_coherence_2022}. Sub-GHz linewidths have been measured in this group of quantum emitters in the absence of applied electric fields or post-processing treatments.

Here we investigate the Stark shift of blue quantum emitters in hBN. We performed coherent excitation to measure the Stark shift upon applying in-plane and out-of-plane electric fields to characterize the electric dipole interaction for blue emitters. Building on these results, we discuss implications for defect symmetry and use density functional theory (DFT) to identify the likely defect structure. Our results constitute an important step towards identifying spectrally-stable quantum emitters in hBN and are particularly valuable for creating site-specific quantum emitters that are resonant with optical cavities or with other quantum emitters.
 
\begin{figure}[b]
\includegraphics[width=1\columnwidth,scale=3]{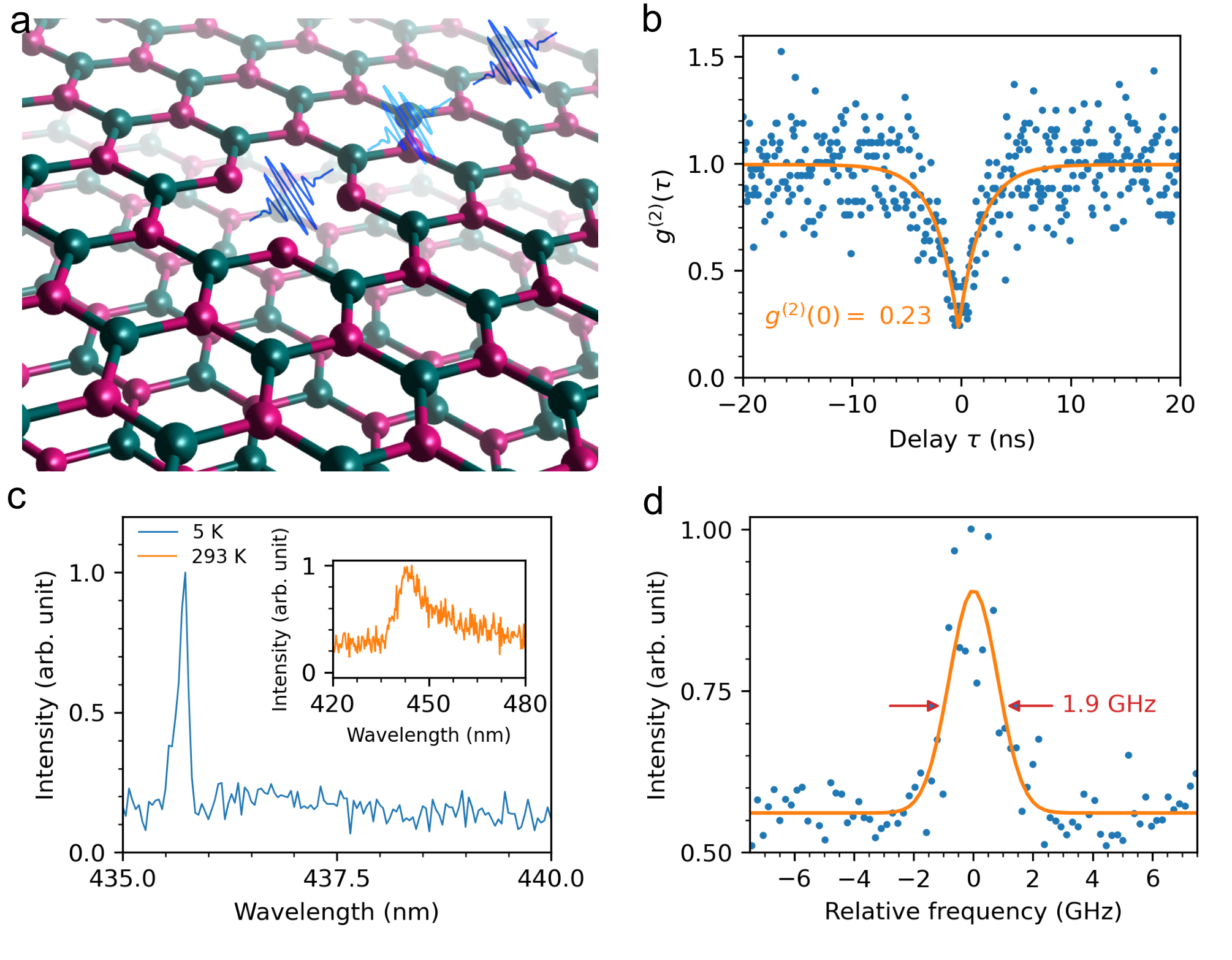}
\caption{Blue single photon emitters in hBN. (a) Atomic defects in hBN as a source of quantum emission. (b) Room temperature autocorrelation measurement of a single defect in hBN emitting at \SI{436}{nm}. The solid line is a fit to the data. (c) Photoluminescence spectra of a blue emitter obtained at cryogenic and room temperatures. (d) Resonant excitation of a blue emitter fit with a Gaussian function that has a line width of \SI{1.9}{GHz}.}
\label{fig:F1}
\end{figure}

In brief, the blue emitters were created by transferring an exfoliated hBN flake onto a silicon substrate with a \SI{285}{nm} thermal oxide, then exposing the flake to a focused electron beam using a commercial scanning electron microscope. Further details on the generation of these emitters in hBN can be found in the Methods section of the Supplementary Information. Figure \ref{fig:F1}(a) shows a schematic illustration of the hBN lattice hosting a point defect, and an autocorrelation measurement confirming the quantum nature of a blue emitter that was fabricated by an electron beam (Fig. \ref{fig:F1}(b)). Whilst the emission energy of single photon emitters in hBN spans a broad spectral range, from UV up to NIR, the zero phonon line (ZPL) of blue emitters is centered consistently on \SI{436}{nm} (Fig. \ref{fig:F1}(c)) and varies by less than \SI{1}{nm} \cite{fournier_position-controlled_2021, horder_coherence_2022}. To measure the linewidth of these emitters, the sample was loaded into a closed-loop cryostat and cooled to \SI{5}{K}. When excited with a \SI{405}{nm} laser, the linewidth of these emitters at cryogenic temperature is beyond the \SI{130}{GHz} resolution of our spectrometer. With a lifetime of approximately \SI{3}{ns}, a natural linewidth of \SI{53}{MHz} is expected from these emitters. By scanning a narrow bandwidth laser across the ZPL, the full width at half maximum of this particular emission is revealed to be \SI{1.9}{GHz}. The resonant photoluminescence excitation (PLE) scan shown in Fig. \ref{fig:F1}(d) is the average of ten individual scans over the ZPL to account for temporal fluctuations of the emission. The best peak fit is Gaussian, indicative of inhomogeneous broadening caused by spectral diffusion. We note that the range of spectral diffusion observed in PLE measurements varied between emitters, but was insufficient to be observed in photoluminescence (PL) spectra which acquired with a spectral resolution of \SI{130}{GHz}. This degree of inhomogeneous broadening is small compared to that of other emitters studied previously in hBN, indicating an inherent insensitivity of the blue emitters to environmental effects such as electric field fluctuations in the hBN lattice.

\begin{figure}[h]
\includegraphics[width=1\columnwidth,scale=3]{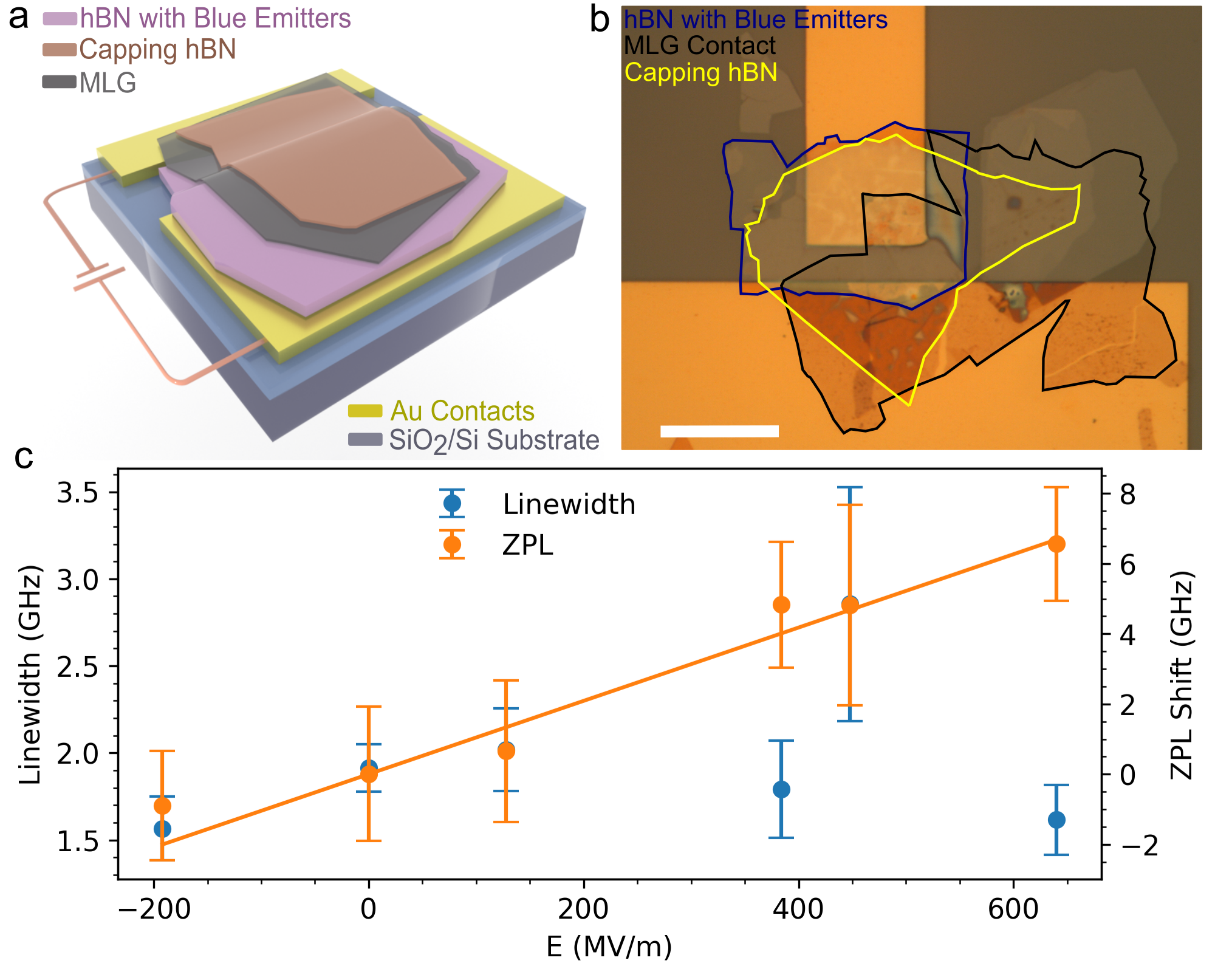}
\caption{Stark effect in a vertical structure used to generate an out-of-plane electric field. (a) Schematic illustration of a vertical structure with hBN located between the bottom gold electrode and a top multilayer graphene (MLG) electrode. (b) Microscope image of the lateral device (the scale bar is \SI{20}{\micro m}). Colored outlines show the edges of each layer: the hBN layer hosting blue emitters is covered with MGL and a protective hBN capping layer. (c) Emission linewidth and ZPL shift versus applied electric field determined by Equ. \ref{equ:E3}. A linear fit yields a gradient of \SI{0.01}{GHz \per MVm^{-1}}.}
\label{fig:F2}
\end{figure}

To characterize the interaction of blue emitters with an external electric field, the Stark effect of these emitters was measured. The transition energy of a color center coupled to an external DC electric field shifts as
\begin{equation}
    \Delta E = - \Delta \mu F   
\label{equ:E1}
\end{equation}
where $\Delta E$ is the Stark shift of the ZPL, $\Delta \mu$ is the difference between the dipole moment of the ground and excited states, and $F$ is the electric field. A nonlinear Stark effect originates from the polarization and hyperpolarization of the defect states that make $\Delta \mu$ dependent on the electric field. In this case, the energy shift $\Delta E$ of the transition line can be expanded as a power series of the electric field \cite{de_santis_investigation_2021, tamarat_stark_2006}
\begin{equation}
    \Delta E = - \Delta \mu ( F)F = - \Delta \mu_{perm} F - \frac{1}{2}\Delta \alpha F^2
\label{equ:E2}
\end{equation}
where $\Delta \mu_{perm}$ is the difference between the permanent dipole of the ground and excited states, and $\Delta \alpha$ is the difference between the polarizability of the ground and excited states. Here the higher order Stark shifts related to hyperpolarizabilities are not considered as their contribution is negligible. Nonlinear electric field dependence is observed when the linear term vanishes, either due to the vanishing permanent dipole moment in both the ground and excited states, or due to the perfect cancellation of the ground and excited state permanent dipoles. The latter is less frequent and appears only in special cases \cite{udvarhelyi_spectrally_2019}.

We fabricated two devices used to apply in-plane and out-of-plane fields to blue emitters. We first built a vertical device (which produces an out-of-plane electric field) by transferring a \SI{36}{nm} hBN flake containing blue emitters over gold electrodes, as shown in Fig. \ref{fig:F2}(a). The electrodes were fabricated using photolithography, followed by vacuum thermal deposition of \SI{5}{nm} Cr then \SI{100}{nm} Au. Multilayer graphene (MLG) was placed on top of the hBN to act as a top electrode, such that we could image the underlying hBN in a confocal microscope. An additional hBN flake was used as a capping layer to cover the MLG and protect it from oxidation during the experiments. The materials were stacked in the manner described in the Methods section of the Supplementary Information, in order to maintain a clean interface. The thickness of each layer was confirmed by AFM. The vertical structure was then irradiated by an electron beam in order to generate blue emitters in the bottom hBN layer. To ensure that emitters were created only in the bottom hBN layer, both hBN flakes were pre-screened prior to stacking for the presence of a signature UV emission, as is detailed in \cite{gale_site-specific_2022}. An optical image of the fabricated vertical device is shown in Fig. \ref{fig:F2}(b). The hBN flake containing blue emitters is highlighted with a blue line around its periphery. The MLG and capping hBN layers are indicated by black and yellow lines, respectively. The device was loaded into a cryostat, and the measured leakage current was limited to few tens of nanoamperes at biases of \SI{\pm 15}{V} applied to the electrodes. 

For the investigated emitter with a linewidth of \SI{1.9}{GHz}, a small linear Stark shift was observed upon application of an out-of-plane electric field by the device. The strength of the applied electric field in the region of the emitter was estimated using the Lorentz local field approximation
\begin{equation}
    F = (\varepsilon_{hBN}+2) \frac{V_{g}}{3d}
\label{equ:E3}
\end{equation}
where $\varepsilon_{hBN} = 3.76$ is the out-of-plane dielectric constant \cite{laturia_dielectric_2018}, $V_g$ is the applied potential field, and $d = 36 \textrm{ nm}$ is the electrode gap, equal to the thickness of the hBN flake containing the blue emitter. A relatively large electric field was required to produce the observed ZPL shift. A linear fit yields a sensitivity of \SI{0.01}{GHz \per MVm^{-1}} or \SI{0.04}{meV \per V nm^{-1}}, indicating that the transition dipole moment of the emitter is either very small or is oriented prediminantly within the plane of the hBN lattice, and hence perpendicular to the electric field. Similarly, the lack of an obvious quadratic dependence suggests that the system has a low transition polarizability in the out-of-plane direction. The emission linewidth was approximately constant, and did not change systematically with applied electric field, confirming that heating caused by the leakage current was negligible. We note that for this particular blue emitter, the emission intensity gradually decreased with increasing negative bias, until the PLE emission was no longer observable which occurred below \SI{-200}{MVm^{-1}}. Similar behavior has been observed in other quantum emitters in hBN and was attributed to charge transfer caused by tuning of the Fermi level \cite{white_electrical_2022, Yu_2022}.

\begin{figure}[b]
\includegraphics[width=1\columnwidth,scale=3]{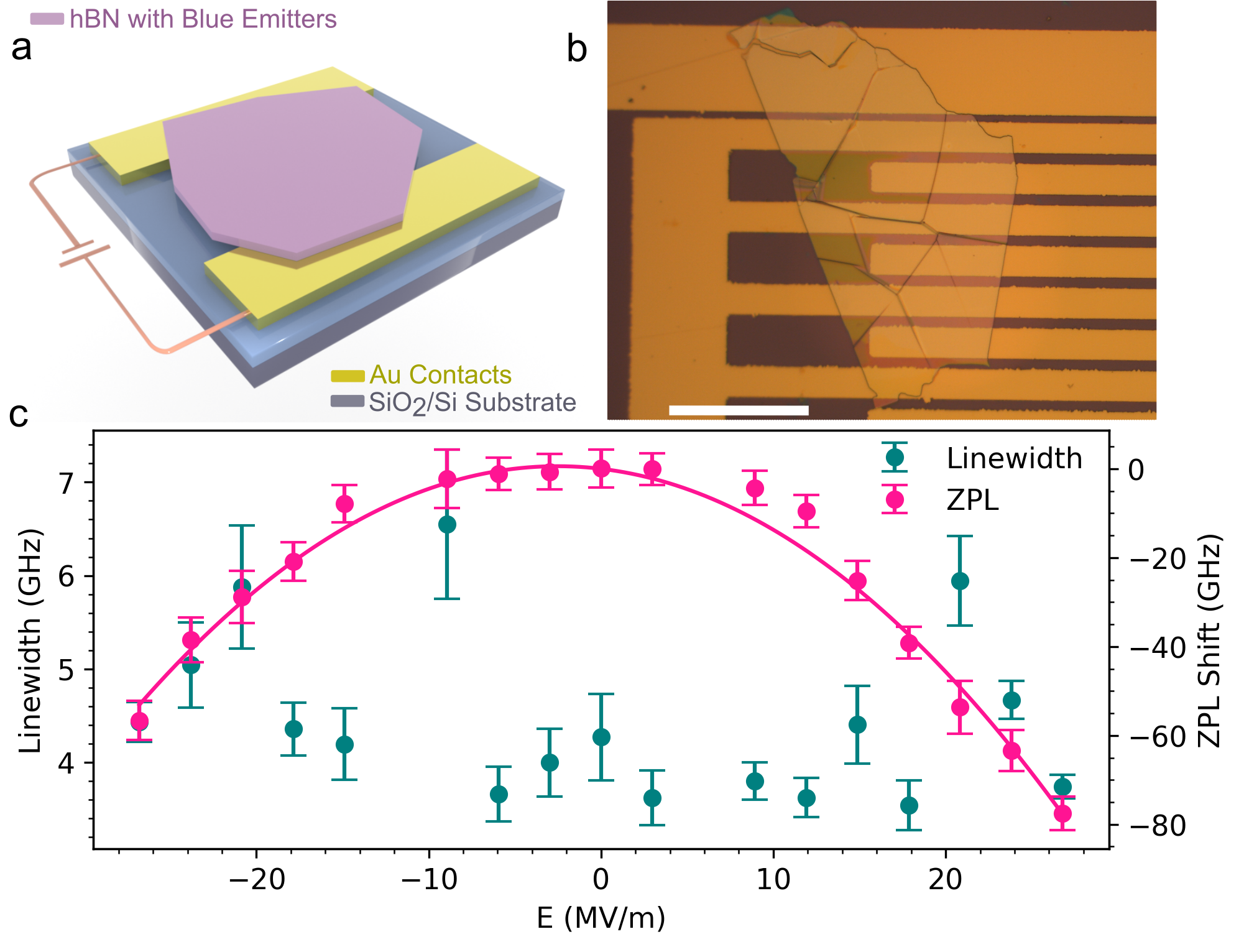}
\caption{Stark effect in a lateral structure used to generate an in-plane electric field. (a) Schematic illustration of a lateral structure with hBN between two gold electrodes. (b) Microscope image of the lateral device (the scale bar is \SI{50}{\micro m}). A hBN flake is positioned on top of interdigitated gold electrodes, with gap separations of \SI{2}{\micro m}. (c) Linewidth and ZPL shifts versus applied electric field, as determined by Equ. \ref{equ:E3}. The ZPL shift data are fit to Equ. \ref{equ:E2}, yielding the values $\Delta \mu = 0.10 \textrm{ D}$ and $\Delta \alpha = 1078 \textrm{ \AA$^3$}$.}
\label{fig:F3}
\end{figure}

We next built a lateral device used to generate an in-plane electric field by transferring a \SI{100}{nm} hBN flake over gold electrodes, as shown in Fig. \ref{fig:F3}(a),(b). The electrode spacing is \SI{2}{\micro m}, which is not large enough for the hBN flake to conform fully to the silicon oxide surface -- the flake remains mostly suspended over the gaps. Nevertheless, the local electric field is assumed to be predominantly in-plane for emitters positioned equidistant between two neighboring electrodes, so that the field strength at the emitter is taken to be equal to the maximum field strength given by the electrode geometry. As with the vertical device, here the local applied electric field is approximated by Equ. \ref{equ:E3}, although it is now appropriate to use the in-plane dielectric constant of $\varepsilon_{hBN} = 6.93$ \cite{laturia_dielectric_2018}, whilst $d = \textrm{\SI{2}{\micro m}}$ is given by the separation of the electrodes. Figure \ref{fig:F3}(b) shows an optical image of the hBN flake overlaying the gold electrodes. The hBN flake was then irradiated by an electron beam to generate blue emitters in the gap between the electrodes. PL characterization of the emitter was performed after confirming that the lateral device did not show any detectable leakage current.

 In contrast to the out-of-plane electric field, the lateral device produced a clear quadratic shift of the emitter ZPL, as shown in Fig. \ref{fig:F3}(c). The shift magnitude per unit electric field strength is over 200 times greater than the corresponding shift in the vertical, in-plane electric field device. A measurement of the emission polarization showed that, for this particular emitter, the light was linearly polarized at an angle of \SI{18}{\degree} with respect to the local applied electric field direction. Assuming that the transition dipole moment is aligned parallel to the polarization of the emitted light \cite{de_santis_investigation_2021}, the projection of the transition dipole moment onto the applied electric field can be determined. Using Equ. \ref{equ:E2} to fit the ZPL data in Fig. \ref{fig:F3}(c), we get the value of $\Delta \mu = 0.10 \textrm{ D}$ for the permanent transition dipole moment, and $\Delta \alpha = 1078 \textrm{ \AA}^3$ for the transition polarizability along the local field direction. A positive transition polarizability indicates that the excited state has greater polarizability than the ground state \cite{tamarat_stark_2006}. These results suggest that the blue emitter has a relatively small permanent transition dipole moment, and a relatively large transition polarizability, relative to visible emitters studied previously in hBN \cite{xia_room-temperature_2019, nikolay_very_2019, noh_stark_2018}.

In the same lateral structure, a nearby emitter was found to also display a small quadratic ZPL wavelength shift in response to the applied electric field. This further suggests that the total transition dipole moment of the blue emitter has a predominant contribution from the transition polarizability, with minimal contribution from the permanent transition dipole moment. Variation in polarizability between emitters has been attributed to both strain \cite{de_santis_investigation_2021} and the dipole orientation within the electric field \cite{tamarat_stark_2006, maze_properties_2011}. Each of these factors can produce a lifting of the degeneracy of the excited state in proportion to its magnitude, leading to a modulated quadratic shift. Emission polarization measurements indicate that these two blue emitters are oriented at \SI{60}{\degree} with respect to one another. This orientation variance, in addition to inherent local strain gradients, could be the cause of the differing quadratic shift.
 
\begin{figure}[b]
\includegraphics[width=1\columnwidth,scale=3]{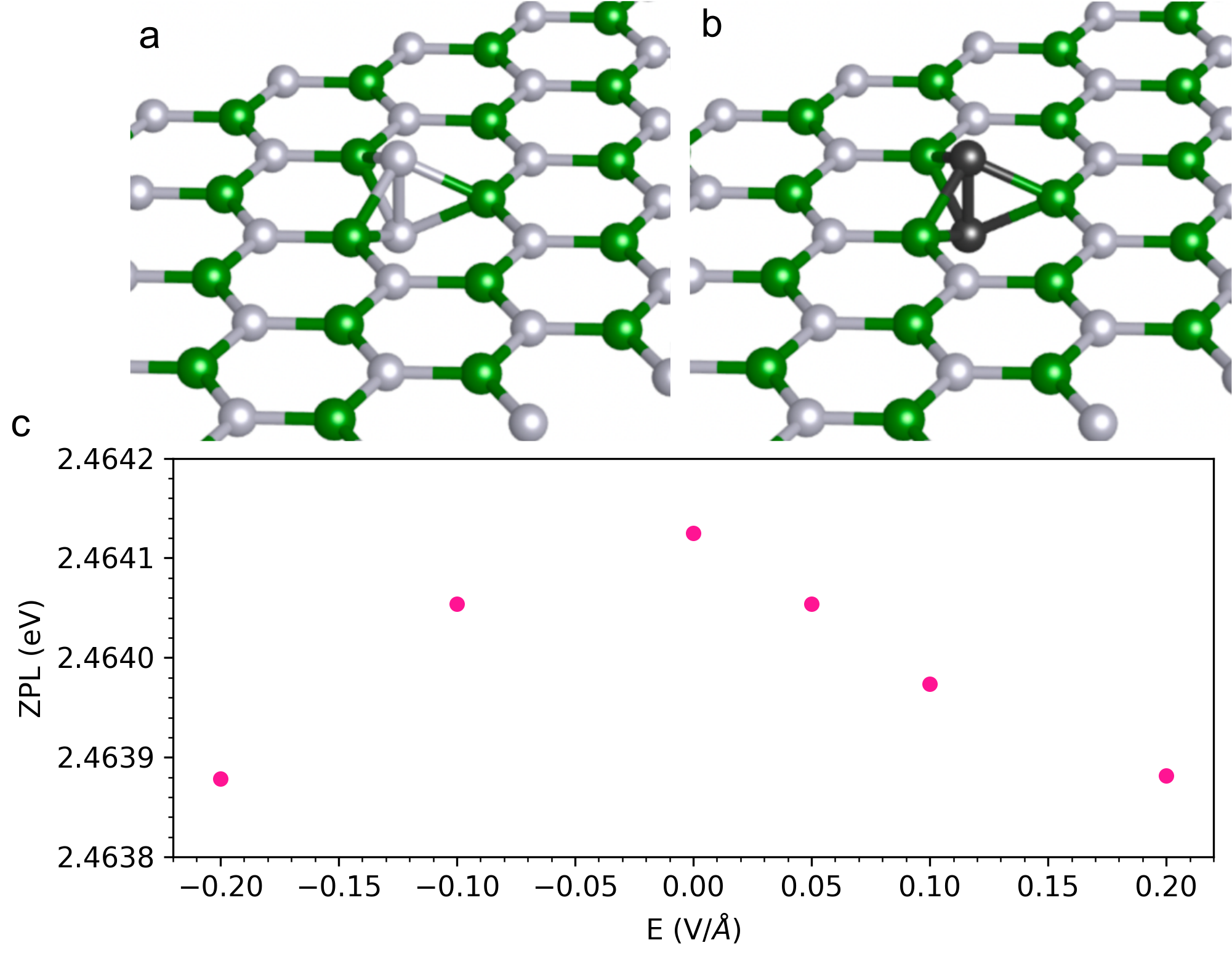}
\caption{Representation of (a) Nitrogen split interstitial defect. (b) $C^2_N$ defect in the hBN lattice. (c) The electric field dependence of the ZPL energy of the nitrogen split interstitial defect as obtained from first principles calculations.}
\label{fig:F4}
\end{figure} 
 
The absence of a permanent dipole can be a consequence of defect symmetry. Indeed, group IV-vacancy complexes in diamond may possess $D_{3d}$ symmetry where the inversion symmetry element of the group ensures a vanishing permanent dipole for these defects. The nonlinear Stark effect has been recently reported for several such complexes in diamond \cite{de_santis_investigation_2021}. The $D_{3d}$ point group symmetry of hBN does not contain inversion symmetry, however, the combination of the 3-fold rotation symmetry and in-plane reflection symmetry may also forbid the presence of a permanent dipole. Using symmetry arguments, the defect that exhibits a nonlinear Stark effect due to the absence of a permanent dipole has to have D3h point group symmetry in both the ground and the excited states.

The observation of a non-linear Stark shift helps us to narrow down the possible point defect configurations that may give rise to the blue PL emission in hBN. The simplest configurations that have $D_{3h}$ symmetry are in-plane single site defects, such as anti-sites, substitutional impurities, and vacancies. Boron and nitrogen vacancies are well known in the literature \cite{ivady_ab_2020, weston_native_2018}, but according to the current understanding of these defects, they are not related to the blue emitter in hBN. Furthermore, we exclude both anti-site and carbon interstitial defects. These defects form strong $sp^2$ hydride bonding states with the neighboring atoms, within the valence band of hBN. The defect states that appear in the band gap for these defects are related to the $p_z$ state of the anti-site or the substitutional carbon. Due to the lack of other defect states in the band gap, the possible optical transitions involve excitations from the valence band to the conduction band of the host.  

For more complex defects the symmetry requirements are fairly restrictive. In-plane two site defects can be ruled out due to the lack of 3-fold rotation symmetry. Consequently, the $C_{B}C_{N}$ defect (associated with the \SI{4.1}{eV} emission) and related defects can be ruled out too.  

Next, we consider interstitial defects. The most favorable configuration of the nitrogen interstitial is a split interstitial configuration \cite{weston_native_2018}, where two out-of-plane nitrogen atoms sit at a nitrogen site, see Fig. \ref{fig:F4}(a). The negative charge state of the defect retains its $D_{3h}$ symmetry in the first optical excited state and possesses no permanent dipole in the ground and excited states. As depicted in Fig. \ref{fig:F4}(c), the ZPL energy calculated on DFT-PBE level of theory depends nonlinearly on the electric field, which is in qualitative agreement with our measurements. However, we note, however, that the blue emitters are generated by electron irradiation of hBN flakes characterized by a \SI{4.1}{eV} UV emission that is associated with carbon \cite{gale_site-specific_2022}. We therefore consider a related atomic structure obtained by replacing the two nitrogen atoms with two carbon atoms as shown in Fig. \ref{fig:F4}(b). We dub this configuration $C_{N}^{2}$. Interestingly, this configuration adopts $D_{3h}$ symmetry, exhibits similar properties of nitrogen split interstitial defect, and can be obtained from $C_{B}C_{N}$ by the addition of a boron interstitial as
\begin{equation}
    C_{B}C_{N}+B_{i} \rightarrow C^{2}_{N} + 2.0\ eV
\label{equ:E4}
\end{equation}
Alternatively, dissociation of the \SI{4.1}{eV} emitter due to the electron irradiation as suggested before \cite{gale_site-specific_2022, horder_coherence_2022} may lead to interstitial carbon atoms which may recombine with carbon substitutional defect and form the $C_{N}^{2}$ complex as
\begin{equation}
    C_{N}+C_{i} \rightarrow C^{2}_{N} + 2.8\ eV
\label{equ:E5}
\end{equation}
The positive energy gain of the reconstruction of the defects ensures the stability of $C_{N}^{2}$ in both processes. Furthermore, since both interstitial boron and carbon atoms are mobile at room temperature \cite{weston_native_2018}, the formation of the $C_{N}^{2}$ defect may happen without any annealing step solely due to electron irradiation.

In conclusion, the emission tunability of blue quantum emitters in hBN has been investigated through the use of two separate devices used to generate in-plane and out-of-plane electric fields. The transition energy is influenced by out-of-plane fields, but the magnitude of the observed linear Stark shift is small relative to the linewidth. In contrast, in-plane electric fields produced significantly greater ZPL modulation, yielding a clear quadratic Stark shift response. Using these results we provide insight into the possible atomic structure of defects responsible for the reported blue emitters in hBN. We suggest the transition dipole moment resides mainly within the plane of the hBN lattice, which may result in decreased sensitivity to local field fluctuations that cause spectral diffusion.

\begin{acknowledgments}
The authors acknowledge financial support from the Australian Research Council (CE200100010) and the Office of Naval Research Global (N62909-22-1-2028) for financial support. The authors thank the UTS node of Optofab ANFF for the assistance with nanofabrication. V.I. acknowledges the support from the Knut and Alice Wallenberg Foundation through WBSQD2 project (Grant No.2018.0071). The calculations were performed on resources provided by the Swedish National Infrastructure for Computing (SNIC) at the National Supercomputer Center (NSC).    
\end{acknowledgments}

\bibliography{Stark-references}

\end{document}